\documentclass[preprint,12pt]{elsarticle}

\usepackage{graphicx}
\usepackage{amssymb}

\usepackage{lineno}

\newcommand{\be}{\begin{equation}}
\newcommand{\ee}{\end{equation}}
\newcommand{\bea}{\begin{eqnarray}}
\newcommand{\eea}{\end{eqnarray}}
\newcommand{\Eq}[1]{Eq.~(\ref{#1})}

\newcommand{\Fig}[1]{Fig.~(\ref{#1})}

\journal{}

\begin{document}

\begin{frontmatter}


\title{Mutual correlation in the shock wave geometry}



\author{ Xiao-Xiong Zeng$^{1}$,Hong-Bo Shao$^{2}$, Ling Li$^{3}$,Xian-Ming Liu$^4$}

\address{$^{1}$School of Material Science and Engineering, Chongqing Jiaotong University,
    Chongqing ~400074, China\\
  $^{2}$ College of Science, Agricultural University of Hebei, Baoding, 071000, China\\
 $^{3}$School of Mathematics, Sichuan  University of Arts and Science, Dazhou~~635000, China\\
 $^{4}$School of Science, Hubei University for Nationalities, Enshi, 445000, China}

\begin{abstract}
We probe the  shock wave geometry with the mutual correlation in a spherically symmetric Reissner Nordstr\"om AdS black hole on the basis of the  gauge/gravity duality. In the static background, we find that the  regions living on the boundary of the  AdS black holes are correlated  provided  the considered regions on the boundary are large enough. We also investigate the effect of the charge on the mutual  correlation and find that the bigger the value of the charge is, the smaller the value of the mutual correlation will to be. As a small perturbation is added at the AdS boundary,  the horizon  shifts and a dynamical shock wave geometry  forms after long time enough.
  In this dynamic background, we find that the greater the shift of the horizon is, the smaller the mutual  correlation will to be.  Especially for the case that the shift is large enough, the mutual correlation  vanishes, which implies that  the considered regions on the boundary are uncorrelated.  The effect of the charge on the mutual  correlation in this dynamic background is  found to be the same as that in the static background.
\end{abstract}


\end{frontmatter}


\section{Introduction}
\label{sec:intro}

Butterfly effect  is   an ubiquitous phenomenon in physical systems.  One progress on this topic recent years is that it also can be addressed in the context of gravity theory \cite{Leichenauer:2014nxa, Berenstein:2015yxu, Sircar:2016old,Ling:2016ibq, Ling:2016ibq2,Shenker11,Shenker12,Shenker13, Fenglu, Alishahiha, Reynolds, Huang2017,Lucas143, Gu129,caizeng} with the help of the AdS/CFT correspondence\cite{ads1,ads2,ads3}. In this framework, one can define the so-called thermofield double state
on the boundary of an  eternal AdS black hole\cite{Maldacena:2013xja}.  As a small perturbation with energy $E$ is added
along the constant $\mu$
 trajectory in the  Kruskal coordinate to one of the boundary at early time
$t_w$, one   find  a bound of infinite energy  accumulates  near the horizon and a shock wave geometry  forms at $t=0$, which is the so-called butterfly effect in the AdS black holes
\cite{Shenker:2013pqa}. The evolution of the shock wave is dual to the evolution of the thermofield double state according to the intercalation of the AdS/CFT correspondence.
The mutual information, defined by
\be\label{mutual}
M(A,B)\equiv S(A)+S(B)-S(A\cup B),
\ee
is often used to probe the effect of the shock wave  on the entanglement of the subsystems $A$ and $B$ living on the boundary \cite{Shenker:2013pqa}, where $S(A)$, $S(B)$ are the entanglement entropy of the space-like regions on $A$ and $B$, which can be calculated by the area  of the minimal surface proposed by Ryu and Takayanagi\cite{23}, while  $S(A\cup B)$ is the entanglement entropy of a  region which cross the horizon and connects $A$ and $B$.

 There are two important quantities characterizing the butterfly effect. One is the scrambling time, which takes the universal form\cite{Shenker:2013pqa}
\be
t_{\star}=\beta \log {S} ,
\ee
where $S$ is the black hole entropy and $\beta$ is the inverse temperature. The scrambling time is the
 time when the mutual information between the two sides on  $A$ and $B$  vanishes. The other is the  Lyapunov exponent $\lambda_L$, which  has the following bound  \cite{Maldacena20162}
 \be
\lambda_L\leq \frac{2 \pi}{\beta},
\ee
  the saturation of this bound has been suggested as
the criterion on whether a many-body system has a holographic dual with a bulk theory\cite{Maldacena20162}. A remarkable
example that saturates this bound is the Sachdev-Ye-
Kitaev  model\cite{Maldacena20162}.

In the initial investigation, the dual black hole geometry  is the non-rotating BTZ black
hole\cite{Shenker:2013pqa}. The area of the  minimal surface equals to the length of the geodesic on the boundary. The mutual information thus is defined by the geodesic length. In this paper, we intend to study butterfly effect in the 4-dimensional  Reissner Nordstr\"om AdS black holes. Though the area of the  minimal surface does not equal to the length of the geodesic, we want to explore whether there is  a quantity defined by the length of the geodesic  can still probe the butterfly effect.  We define this quantity  as  mutual correlation
\be
I(A,B)\equiv L(A)+L(B)-L(A\cup B),
\ee
in which $A$ and $B$ are two points on the left and right boundaries, $L(A)$, $L(B)$ are the space-like geodesic that go through points $A$ and $B$ respectively, and $L(A\cup B)$  is the geodesic length cross the horizon and connects  $A$ and $B$.
 The results are not expectable since
 we can not view simply the mutual correlation  as the spatial section of the mutual information  by fixing some of the transverse coordinates. The metric components of the transverse coordinates are not one  but the  functions of the radial coordinate $r$  so that they have contributions  to the area of the minimal surface.

In the 4-dimensional spacetimes, though the geodesic length does not equal to the area of the minimal surface, it has been shown that both  the geodesic length and area of the minimal surface, which are dual to the two point correlation function and entanglement entropy respectively, are nonlocal probes and have the same effect as they are used to probe the thermalization behavior and phase transition process\cite{1,2,3,4,5,6,7,8,9,10,11,12,13,14,15,16,17,18}. Thus it is interesting to explore whether the
 mutual correlation can probe the butterfly effect as the mutual information for both of them are defined by the nonlocal probes.

In \cite{Leichenauer:2014nxa}, the author has  probed the shock wave geometry  with   mutual information in the 4-dimensional plane symmetric  Reissner Nordstr\"om AdS black branes. They have obtained some analytical results approximately and found that
for large regions the mutual information is positive   in the static black hole,  and  the mutual information will be disrupted as a small perturbation is added in dynamic background.
 In this paper, we  will employ the mutual correlation   to probe the shock wave geometry in the 4-dimensional spherically symmetric  Reissner Nordstr\"om AdS  black holes. Our motivation is twofold. On one hand, we intend to give the exact numeric result between  the size of the boundary region and mutual correlation as  well as the perturbation  and mutual correlation. One the other hand, we intend to explore how the charge affects the mutual correlation in cases without and with a perturbation. Both cases have not been reported previously in \cite{Leichenauer:2014nxa}.

Our paper is outlined as follows.
 In
sect. 1, we will construct the shock wave geometry  in  the Reissner Nordstr\"om AdS black holes. In sect. 2, we will study  the mutual correlation in the static background. We concentrate on the effect of the boundary separation and charge on the mutual correlation.
 In sect. 3, we will probe the  butterfly effect with the  mutual correlation in the dynamical background. We concentrate on  studying the effect of the perturbation and charge
  on the mutual  correlation.
The conclusion and discussion is presented  in sect. 4.
Hereafter in this paper we use natural units ($G=c=\hbar=1$) for
simplicity.

\section{Shock wave geometry  in the Reissner Nordstr\"om AdS black holes}

Starting from the  action,
\begin{eqnarray}\label{rnm}
S&= -\frac{1}{16\pi G}\int d^{d+1}x \sqrt{g}\left(\mathcal{R} + \frac{d(d-1)}{\ell^2} - \frac{1}{4} F_{\mu\nu}F^{\mu\nu}\right),
\end{eqnarray}
one can get the Reissner-Nordstr\"om AdS black holes solution. For the case $d=3$, we have
\begin{eqnarray}\label{peturbation}
ds^{2}=-f(r)dt^2+\frac{dr^2}{f(r)}+r^{2}(d\theta^2+\sin^2\theta \phi^2),
\end{eqnarray}
in which $f(r)=1-\frac{2M}{r}+\frac{Q^2}{r^2}+r^2$,
where  $M$ is the mass and $Q$ is the charge of the black hole.

In order to discuss the  butterfly effect of a black hole, one should construct the shock wave geometry in the Kruskal coordinate firstly.
We will review the key procedures and give the main results as done in\cite{Shenker:2013pqa} for the consistency of this paper though there have been some discussions on this topic.

 The event horizon, $r_h$, of the black hole is determined by $f(r_h)=0$. With the definition  of the surface gravity, $\kappa=f(r)^{\prime}\mid_{r_h}/2$, we also can get the Hawking temperature $T=\kappa/2\pi$, which is regarded as the temperature of the dual conformal field theory  according to the AdS/CFT correspondence.
In the  Kruskal coordinate system, the metric in Eq.(\ref{peturbation}) can be rewritten as
\begin{eqnarray}\label{kruskal}
ds^{2}=\frac{1}{\kappa^2} \frac{f(r)}{\mu\nu}d \mu d\nu+r^{2}(d\theta^2+\sin^2\theta \phi^2),
\end{eqnarray}
in which
\begin{eqnarray}
\mu=\pm e^{-\kappa U}, \nu=\mp e^{\kappa V},
\end{eqnarray}
\begin{eqnarray} \label{uv}
\mu\nu=-e^{2\kappa r_{\star}}, \mu/\nu=-e^{-2\kappa t},\label{mu}
\end{eqnarray}
where $U=t-r_{\star}$,  $V=t+r_{\star}$, are the Eddington  coordinate, which are defined by the tortoise coordinate $r_{\star}=\int \frac{dr}{f(r)}$. We will suppose $\mu<0,\nu>0$ at the right exterior as in\cite{Shenker:2013pqa}.  As $r$ approaches to the event horizon and boundary, we know  $r_{\star}$ approaches to $-\infty$ and 0 respectively. Thus from Eq.(\ref{mu}), we know that the event horizon and boundary locate at $\mu\nu=0$ and  $\mu\nu=-1$ respectively.

\begin{figure}[h]
\centering
\includegraphics[scale=0.38]{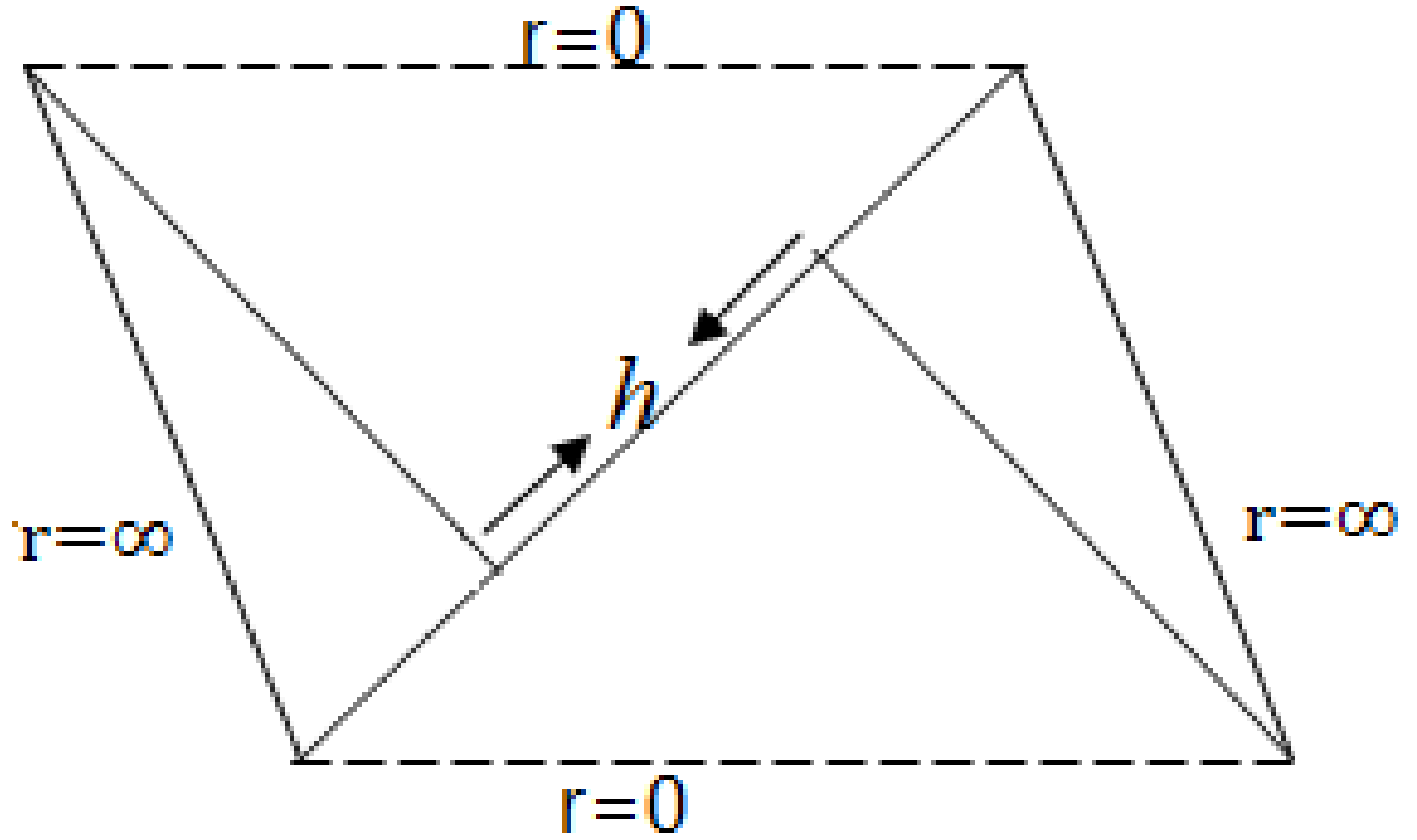}
 \caption{\small Penrose diagrams for an eternal black hole with a  perturbation.} \label{fig1}
\end{figure}

Next we will check how the spacetime changes as a small perturbation with asymptotic energy $E$ is added on the left boundary at time $t_w$ follows  a constant $\mu$ trajectory. We label the  Kruskal coordinate on the left side and right side as $\mu_L, \nu_L$ and  $\mu_R, \nu_R$. The constant $\mu$ trajectory propagation  of the perturbations implies
\begin{eqnarray}
\mu_L=\mu_R=e^{-\kappa t_w}.
\end{eqnarray}
To find the relation between $\nu_L$ and $\nu_R$, we will employ the relation
\begin{eqnarray}
\mu_L\nu_L=-e^{2\kappa_L r_{ \star L}}, \mu_R\nu_R=-e^{2\kappa_R r_{ \star R}}.
\end{eqnarray}
Generally speaking, $\kappa_L=\kappa_R=\kappa$ for the energy $E$ of the perturbation is  much smaller than that of the black hole mass $M$.  On the other hand, we are interested in the case $t_w\rightarrow \infty$, which implies $r\rightarrow r_h$. In this case, we can approximate $ r_{ \star}\approx\frac{1}{2 \kappa}(\log^{r-r_h}+c)$ for there is a relation $f(r)=f^{\prime}(r_h)(r-r_h)+\cdots$. In this case, $e^{2 \kappa r_{ \star}}=C(r-r_h)$, here $C=e^c$. So we have the identification
\be \label{vl}
v_L = v_R +C e^{\kappa t_w} (r_{hL}-r_{hR})  \equiv v_R + h,
\ee
where we have used the relation $C_L=C_R=C$.
From \Eq{vl}, we know that there is a shift in the Kruskal coordinate $\nu$  as the small perturbation across the $\mu=0$  horizon of the black hole. For computations, the shift in $\nu$  is often written as $\nu\rightarrow \nu+h(\theta) \Theta(\mu)$, where $\Theta(\mu)$ is  a step function.
In this case, the \Eq{kruskal} changes into a standard shock wave
\begin{eqnarray}
ds^{2}=A(\mu\nu) d \mu d\nu-A(\mu\nu)h(\theta)\delta (\mu) d\mu^2
+B(\mu\nu)(d\theta^2+\sin^2\theta \phi^2),
\end{eqnarray}
in which we have used the relation $\Theta(\mu)^{\prime}=\delta (\mu)$ and the replacement
\begin{eqnarray}
A(\mu\nu)=\frac{1}{\kappa^2} \frac{f(r((\mu\nu))}{\mu\nu},
B(\mu\nu)=r(\mu\nu)^{2}.
\end{eqnarray}
The Kruskal diagram for the perturbed space time is shown in  \Fig{fig1}.

\section{Mutual correlation in the static Reissner Nordstr\"om AdS black holes}

In this section, we will investigate the mutual correlation in the static background. Our objective is to
explore whether the boundary regions of the AdS black holes are correlated so that we can investigate the effect of the shock wave on the mutual correlation in the next section.

As  depicted in \Fig{fig1}, an eternal black hole has  two asymptotically AdS regions, which  can be holographically described by
two identical, non-interacting copies of the conformal field theory. One thus can define the so-called thermal double  state and study their entanglement and correlation. Our objective is to compute the mutual correlation of a  point $A $ on the left asymptotic boundary and its partner $B$ on the right asymptotic boundary. We will let $A=B$  so that the left and right boundaries are identical.
For the  spherically symmetric black holes in this paper, the AdS boundary   is a 2-dimensional sphere with finite volume. In light of the symmetry of $\phi$ direction, we will use $\theta$ to parameterize the geodesic length between any two points on the boundary, named as $\theta_1,
\theta_2$.

On the left boundary, the geodesic length that go through point A with
boundary separation  $\theta_0$ is
\be \label{false}
L_A=\int dS=\int d\theta\sqrt{f^{-1}r'^2+r^2},
\ee
where $r'=dr/d\theta$. If regarding the integrand in \Eq{false} as the Lagrangian, we can define
a conserved quantity associated with translations in $\theta$, that is
\be \label{conse}
\frac{r^2}{\sqrt{r^2+f^{-1}r'^2}}=r_{\rm min},
\ee
where $r_{\rm min}$ is the turning point of the surface where $dr/d\theta = (\theta')^{-1} = 0$. According to the symmetry, it locates at $\theta=\theta_0/2$. With Eq.(\ref{conse}), $\theta_0$ can be written as
\be \label{theta0}
\theta_0 = \int d\theta= 2\int_{r_{\rm min}}^{\infty} \frac{dr}{r\sqrt{f}}\, \frac{1}{\sqrt{\left(r/r_{\rm min}\right)^{2}- 1}}.
\ee
The geodesic length also can be rewritten as
\be
L_A = 2 \int_{r_{\rm min}}^\infty dr~\frac{1}{\sqrt{f}}\frac{1}{\sqrt{1-(r_{\rm min}/r)^{2}}}.
\ee
Since $B$ is  identified with $A$, $L_B$ thus takes the same form as $L_A$ provided the two points on the boundary located at the same place. As stressed in the introduction, we will employ the mutual correlation to study the correlation
between points $A$ and $B$. Thus our next step is to find $L_{A\cup B}$, which is the
geodesic length connected the left point and right point by passing through
the horizon of the black hole,
 where $\theta'=0$. The total length, including both sides of the horizon,  can be expressed as
\be
L_{A\cup B} =4 \int_{r_{h}}^{\infty} dr~\sqrt{f^{-1}}.
\ee
Putting all these results together, the mutual correlation can be expressed as
\be \label{totallength}
I(\theta_0)=4 \int_{r_{\rm min}}^\infty dr~\frac{1}{\sqrt{f}}\frac{1}{\sqrt{1-(r_{\rm min}/r)^{2}}}-4 \int_{r_{h}}^{\infty} dr~\frac{1}{\sqrt{f}}.
\ee
\begin{figure}[h]
\centering
\includegraphics[scale=0.65]{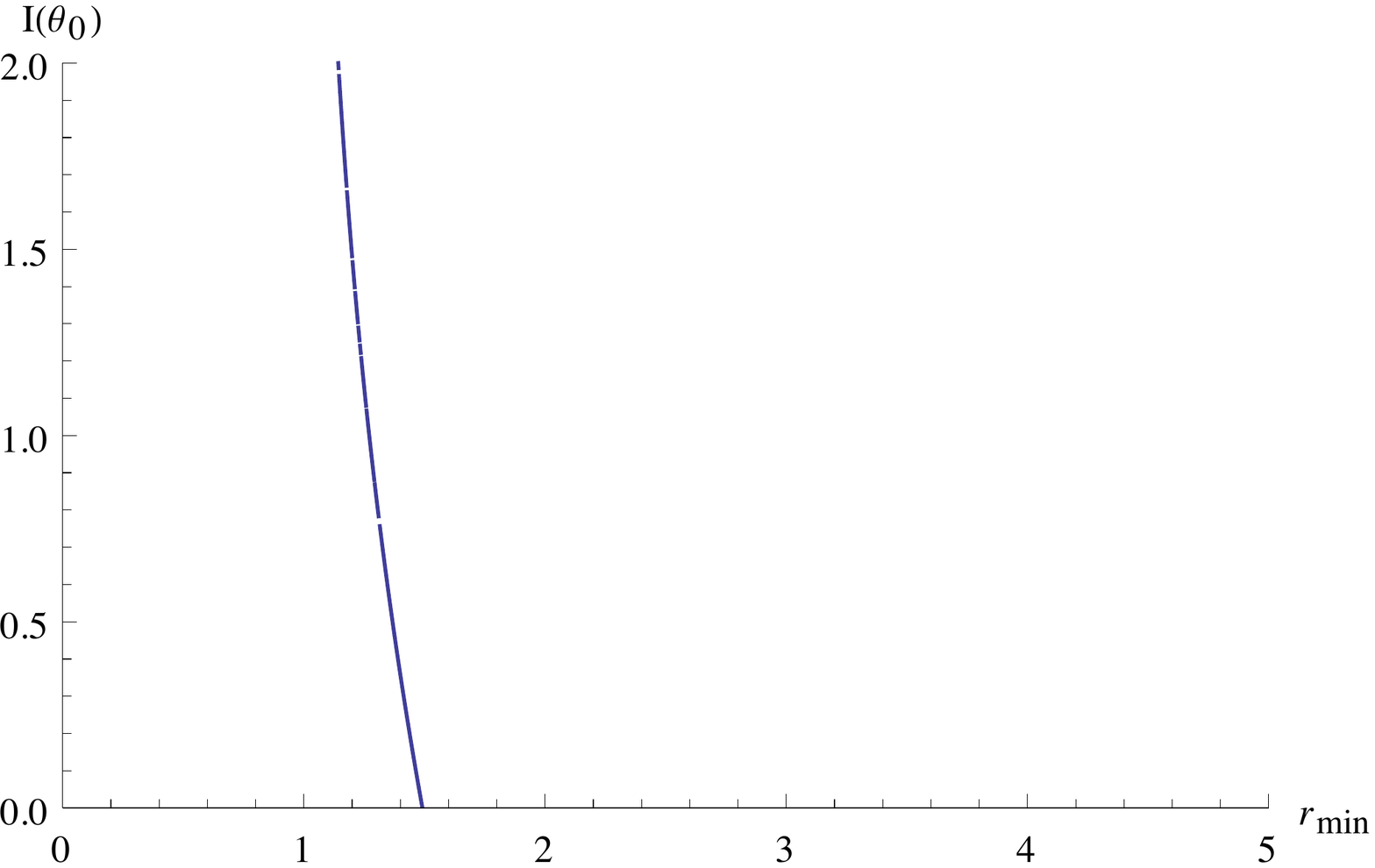}
 \caption{\small  Relation between  $I(\theta_0)$ and $r_{min}$  for the case $Q=0.5$.  } \label{fig31}
\end{figure}
 From Figure \Fig{fig31}, one can read off the relation between the  mutual correlation  and the position of the turning point $r_{\rm min}$. From this figure,  we know that  $I(\theta_0)$ decreases as the value of  $r_{min}$ becomes smaller, and $I(\theta_0)$ vanishes as  $r_{min}$ is larger than $ r_h$ a little. Especially,  as $r_{\rm min}\to r_h$ the mutual correlation will diverge.  That is to say, $r_{\rm min}$ can not penetrate into the black hole, which  was also  observed in \cite{Hubeny8}  where the properties of the geodesic length has been investigated extensively.

\begin{figure}[h]
\centering
\includegraphics[scale=0.65]{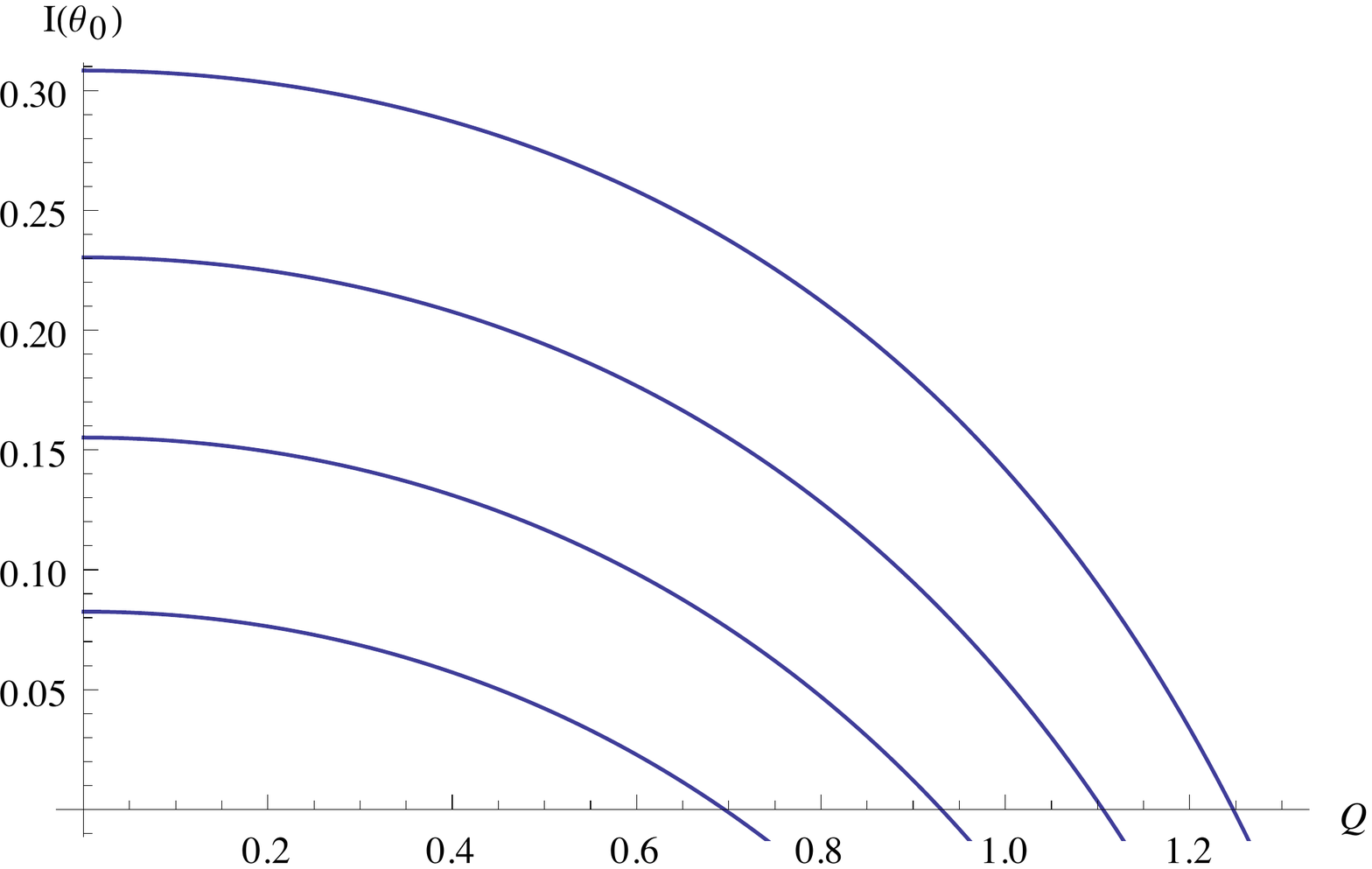}
 \caption{\small   Relation between  $I(\theta_0)$ and $Q$. Curves from top to down represent $r_{min}$ increases from 1.42 to 1.48 with step 0.02. For both cases, we have set $r_h$=1.  } \label{fig32}
\end{figure}

We also can study the effect of  $Q$ on the mutual  correlation $I(\theta_0)$, which is shown in Fig.\ref{fig32}.
 From this figure, we know that $I(\theta_0)$ decreases as  $Q$ grows for a fixed $r_{min}$. There is also a critical charge $Q_c$ where the mutual correlation vanishes, which means that there is no correlation  between the paired subregions we considered. For different $r_{min}$, the value of the critical charge is different. As $r_{min}$ increases, the value of the critical charge decreases.
  For a   fixed $Q$, we find that the mutual correlation is smaller for greater $r_{\rm min}$.

We are interested in how the boundary separation $\theta_0$ affects the mutual correlation, especially to each extent, the mutual correlation vanishes. We thus should express the mutual correlation as a function of  the boundary separation. Substituting \Eq{theta0} into \Eq{totallength}, we obtain
\be \label{totallength1}
I(\theta_0)=2 \theta_0 r_{min}+4 \int_{r_{\rm min}}^\infty dr~\frac{1}{\sqrt{f}}\sqrt{1-(r_{\rm min}/r)^{2}}-4 \int_{r_{h}}^{\infty} dr~\frac{1}{\sqrt{f}}.
\ee

\begin{figure}[h]
\centering
\includegraphics[scale=0.65]{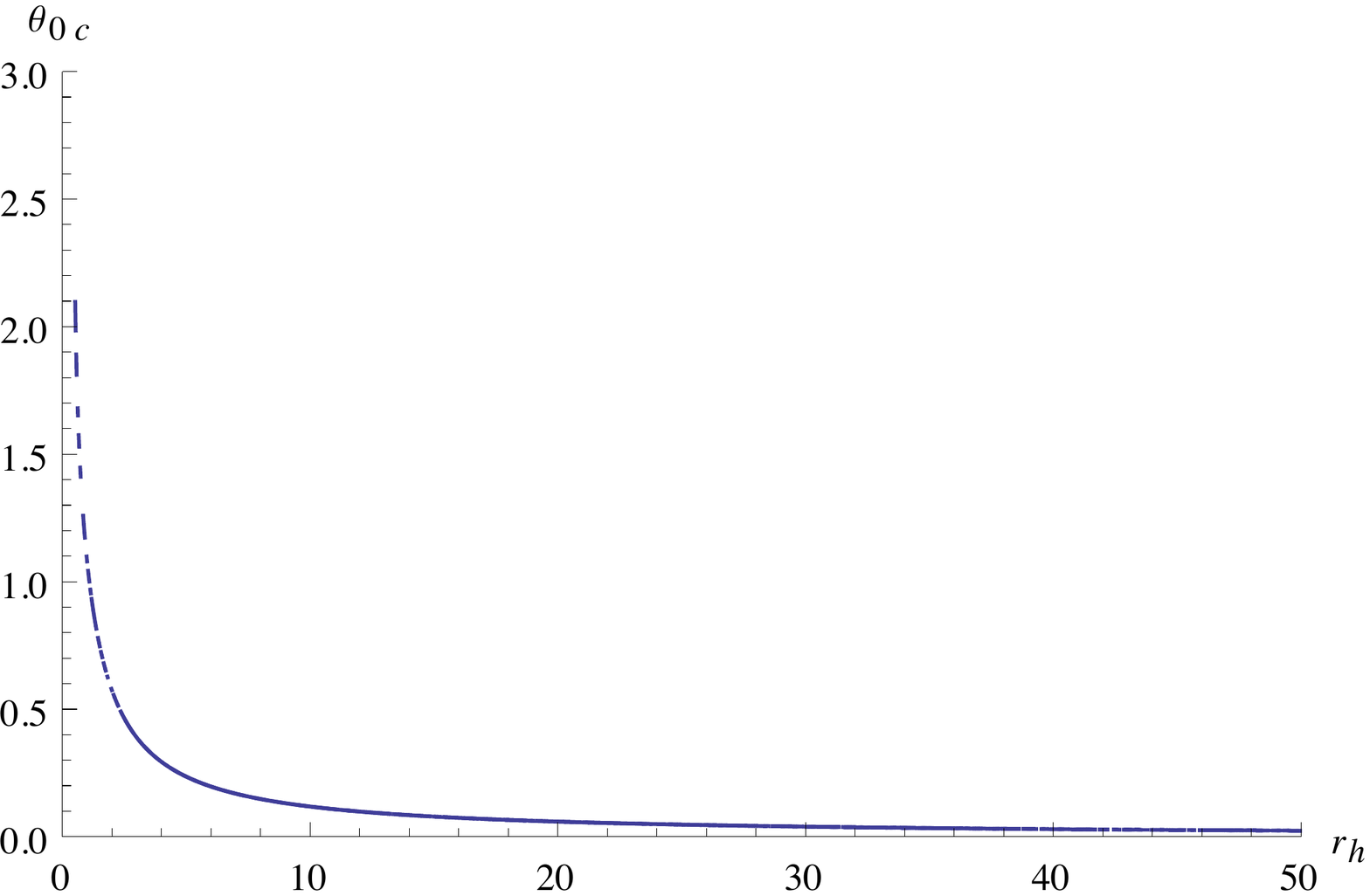}
 \caption{\small Relation between  $\theta_{0c}$ and $r_{h}$ for  the case $Q$=0.5. } \label{fig41}
\end{figure}

\begin{figure}[h]
\centering
\includegraphics[scale=0.65]{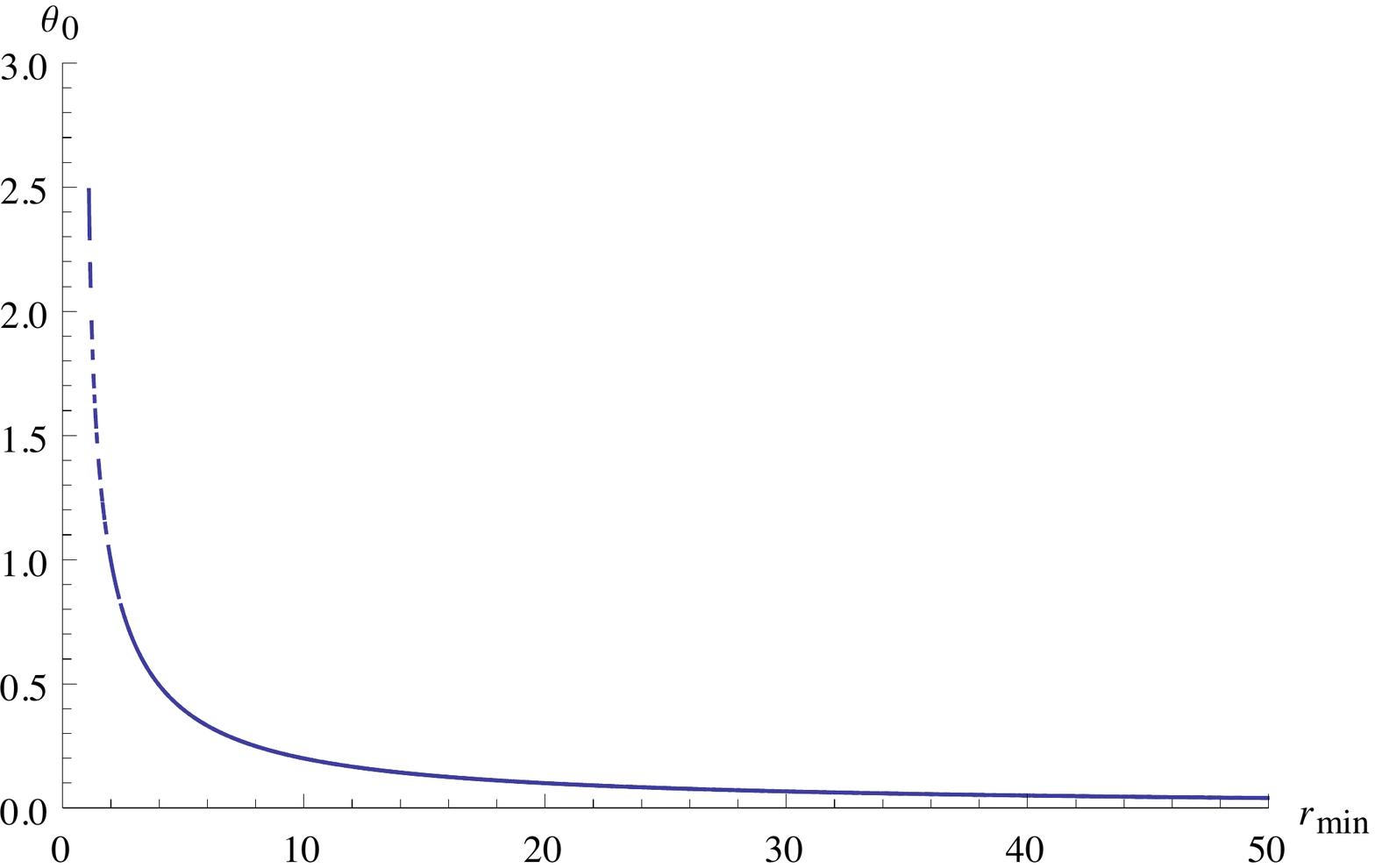}
 \caption{\small Relation between  $\theta_0$ and  $r_{min}$ for  the case $Q$=0.5.} \label{fig42}
\end{figure}

From  Fig.\ref{fig31}, we know that $I(\theta_0)$ will vanish as $r_{min}\simeq r_h$. With this approximation,   the critical value of the boundary separation  in  \Eq{totallength1} can be expressed as
\be \label{totallength2}
\theta_{0c}=\frac{2}{r_{h}}[ \int_{r_{h}}^{\infty} dr~\frac{1}{\sqrt{f}}(1-\sqrt{1-(r_{h}/r)^{2}})].
\ee
With \Eq{totallength2}, we  can discuss how the
 critical value of the boundary separation  $\theta_{0c}$ changes with respect to the horizon $r_h$. From \Fig{fig41}, we know that  $\theta_{0c}$ decreases as  $r_h$ increases. For large enough $r_h$,  $\theta_{0c}$  vanishes. In the  small $r_h$ region,  $\theta_{0c}$ changes sharply as $r_h$  increases. \Fig{fig42} is helpful for us to understand \Fig{fig41}. As we addressed previously,  $\theta_{0c}$  is obtained at $r_h\thickapprox r_{min}$. The relation between  $\theta_{0c}$ and $r_h$ thus is similar to that of $\theta_{0}$ and $r_{min}$. As $r_{min}= \infty$, the geodesic length, and further  the boundary separation,  approach to  zero naturally.

 We already know that bigger $r_{\rm min}$ actually corresponds to smaller separation on the boundary. Therefore, Fig.\ref{fig32} also indicates that smaller subregions have smaller mutual correlation between them, which is consistent with the physical intuition.

\section{Probe the shock wave geometry  via mutual correlation}

\begin{figure} [h]
\centering
\includegraphics[scale=0.42]{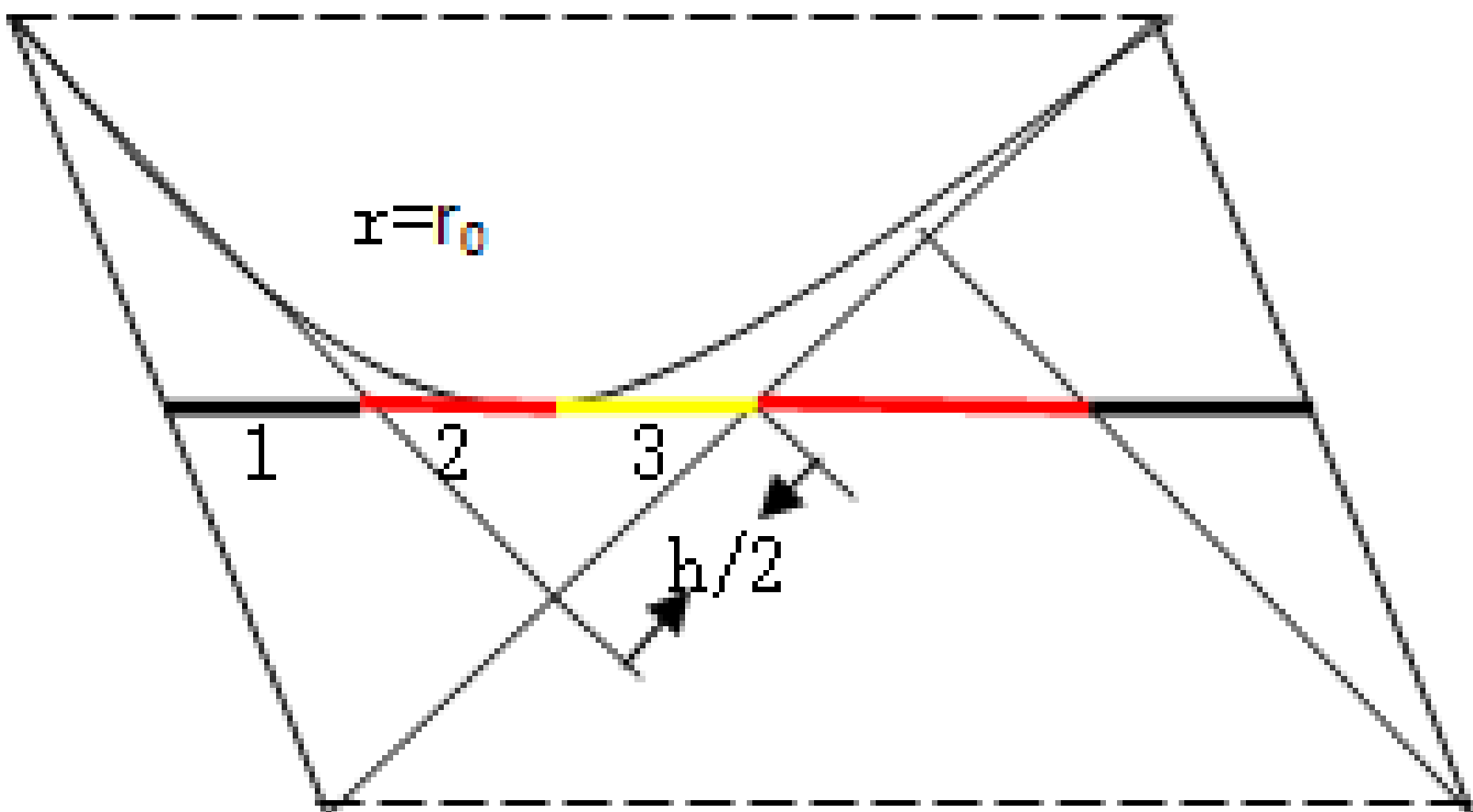}
 \caption{\small  The Penrose diagram and geodesic length (horizontal colourful line) in the shock wave geometry. The left half of the surface is divided  into three segments, labeled by black line, red line and yellow line. The smallest value of $r$ attained by the surface is $r=r_0$, which marks the division between 2 and 3.} \label{fig9}
\end{figure}

As a small perturbation is added from the left boundary,  there is a shift in the $\nu$
 direction for enough long time $t_w$. A shock wave geometry forms and the  passage  connected the left region and right region, namely the wormhole,   is disrupted. In this section,  we intend to investigate the effect of  the disrupted geometry on the mutual correlation.  As in section 3, we suppose point $A$  belongs to the left asymptotic boundary and its identical partner $B$ belongs to the right asymptotic boundary.
At $t=0$, the geodesic length $L_A$ and $L_B$ are unaffected by the shock wave because they do not cross the horizon. However,  the quantity $L_{A\cup B}$ will be affected by the shock wave for it stretches across the wormhole, which is shown in Fig.\ref{fig9}.

 In light of the identification between $A$  and  $B$ as well as the symmetry of the transverse space, we only should calculate the geodesic length for the region 1, 2 and 3 in Fig.\ref{fig9} for the length of the other part is the same as this part.  At a constant $\theta$ surface, the induced metric   can be written as
 \begin{eqnarray}
dx^{2}=[-f(r)+\frac{1}{f(r)}\dot{r}^2]dt^2+r^{2}\sin^2\theta \phi^2,
\end{eqnarray}
 in which we have used $r$ to parameterize the surface and $\dot{r}=dr/dt$. The  geodesic length for the region 1, 2 and 3 in \Fig{fig9}  is then given by
\be \label{ab}
\bar{L}_{A\cup B}(h) = \int dt~\sqrt{-f  +f^{-1}\dot{r}^2}.
\ee

It should be stressed that in \Fig{fig9}, the boundary is a 2-dimensional spherical surface in the Penrose diagram strictly. In this paper, we only consider the geodesic length and neglect the contribution of the $\phi$ direction.

If regarding the integrand  in \Eq{ab} as the Lagrangian, we can define the `Hamiltonian' $\mathcal{H}$ as
\be \label{aab}
\mathcal{H}= \frac{-f }{\sqrt{-f + f^{-1}\dot{r}^2}} =  \sqrt{-f_0},
\ee
in which $f_0= f(r_0)$ and  $r_0$  is  the radial position behind the horizon that satisfies $\dot{r}=0$. From  \Eq{aab}, we know that as  $r_0\to r_h$,  $\mathcal{H}\to 0$,  which correspond to the case that the shock wave is absent for $h \rightarrow 0$ in this case.
With the conservation equation, the $t$ coordinate can be written as a function of $r$
\be\label{eqtime}
t(r) =\pm \int \frac{dr}{f\sqrt{1 +\mathcal{H}^{-2}f}},
\ee
where $\pm$  denote  $\dot{r}>0$ and $\dot{r}<0$ respectively.  Substituting \Eq{eqtime} into \Eq{ab}, we can get a time independent integrand
\be \label{}
\bar{L}_{A\cup B}(h) = \int dr~\frac{1}{\sqrt{\mathcal{H}^2+f}}.
\ee
With this relation, we will compute the geodesic length  starts at $t=0$ on the left asymptotic boundary and ends at $\nu=h/2$ on the horizon, namely the geodesics length of region 1+2+3 in \Fig{fig9}, which can be expressed as
\be \label{}
\bar{L}_{A\cup B}(h) = \int_{r_h}^{\infty} dr~\frac{1}{\sqrt{\mathcal{H}^2+f}}+2 \int_{r_0}^{r_h} dr~\frac{1}{\sqrt{\mathcal{H}^2+f}}.
\ee
The second term contains a prefactor 2 stems from the fact that  the second and third segments in  \Fig{fig9} have the same length.   The total geodesic length, defined as $L_{A\cup B}(h)$,  connected the left boundary and right boundary thus is
\be \label{}
L_{A\cup B}(h) = 2\int_{r_h}^{\infty} dr~\frac{1}{\sqrt{\mathcal{H}^2+f}}+4 \int_{r_0}^{r_h} dr~\frac{1}{\sqrt{\mathcal{H}^2+f}}.
\ee
It should be stressed that the first segment contains a divergent $h$-independent contribution which must be subtracted as we study it numerically. Considering the contribution of $L_A$ and  $L_B$, the mutual correlation  in the shock wave geometry can be expressed as
\be \label{ih}
I(h, \theta_0) = 4 \int_{r_{\rm min}}^{\infty}dr~\frac{1}{\sqrt{f}}\frac{1}{\sqrt{1-(r_{\rm min}/r)^{2}}}-2\int_{r_h}^{\infty} dr~\frac{1}{\sqrt{\mathcal{H}^2+f}}-4 \int_{r_0}^{r_h} dr~\frac{1}{\sqrt{\mathcal{H}^2+f}}.
\ee
Of course, the first term on the right is divergent on the boundary, the contribution from the pure AdS  should be subtracted as we calculate it numerically.

For a fixed $r_h$, we know that $I(h, \theta_0)$ depends on the location of $r_0$. The main objective of this section is to probe the shock wave geometry with the mutual correlation, we thus should find the relation between  $I(h, \theta_0)$ and $h$. To proceed, we should find the relation between  $h$ and $r_0$.

 Firstly, we should find the coordinates of the three segments in \Fig{fig9}.  The first segment goes from the boundary at $(\mu,\nu) = (1,-1)$ to $(\mu,\nu) = (\mu_1,0)$, in which
\be
\mu_1 =  \exp[-\kappa \int_{r_{h}}^{\infty} \frac{dr}{f} (1- \frac{1}{\sqrt{1 +\mathcal{H}^{-2}f}})],
\ee
where we have used \Eq{uv}. The second segment stretches from $(\mu_1,0)$ to  $(\mu_2,\nu_2)$ at which $r = r_0$. The coordinate $\mu_2$ can be determined by the relation
\be
\frac{\mu_2}{\mu_1} =  \exp[-\kappa \int_{r_{0}}^{r_h} \frac{dr}{f} (1- \frac{1}{\sqrt{1 +\mathcal{H}^{-2}f}})].
\ee
 The coordinate $\nu_2$ can be determined by choosing a reference surface $r=\bar{r}$ for  which $r_{\star}=0$ in the black hole  interior. In this case,
\be
\nu_2 = \frac{1}{\mu_2} \exp(2 \kappa \int_{\bar{r}}^{r_0} \frac{dr}{f}).
\ee
The third segment stretches from $(\mu_2,\nu_2)$ to  $(\mu_3=0,\nu_3=h/2)$. With the relation
\be
\frac{\nu_3}{\nu_2} = \frac{h}{2 \nu_2} =\exp[\kappa \int_{r_{0}}^{r_h} \frac{dr}{f} (1- \frac{1}{\sqrt{1 +\mathcal{H}^{-2}f}})]=\frac{\mu_1}{\mu_2},
\ee
we can express $h$ as
\be \label{h}
h=2 \exp(\Pi_1+\Pi_2+\Pi_3),
\ee
where
\be
\Pi_1 = 2 \kappa \int_{\bar{r}}^{r_0} \frac{dr}{f},
\ee

\be
\Pi_2=2 \kappa \int_{r_{0}}^{r_h} \frac{dr}{f} (1- \frac{1}{\sqrt{1 +\mathcal{H}^{-2}f}}),
\ee
\be
\Pi_3= \kappa \int^{\infty}_{r_h} \frac{dr}{f} (1- \frac{1}{\sqrt{1 +\mathcal{H}^{-2}f}}).
\ee

\begin{figure} [h]
\centering
\includegraphics[scale=0.65]{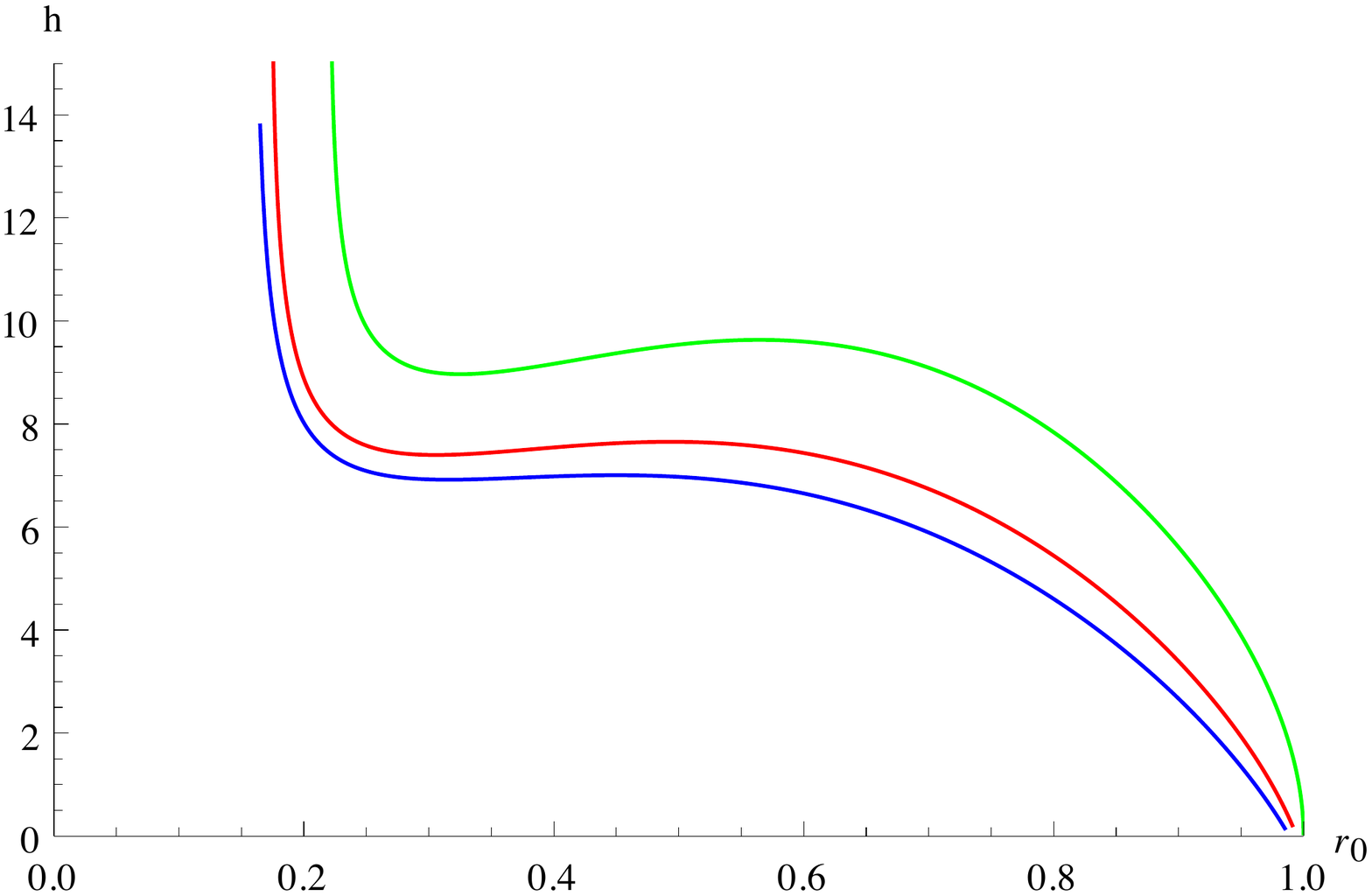}
 \caption{\small  Relation between  $h$ and $r_0$ for the case $ \bar{r}=0.2, r_h=1$. The green line, red line, and blue line correspond to $Q=0.5, 0.52, 0.54$ respectively.  } \label{fig101}
\end{figure}
It is obvious that $h$ depends on the location of $r_0$ for a fixed $r_h$. The relation between $I(h, \theta_0)$ and $h$  is shown in \Fig{fig101}. From this figure, we can see that for a fixed charge  the  relation between  $r_0$  and  $h$ is nonmonotonic. Here we are interested in two locations  on the horizontal axis. One is the initial location of the curve  where $h$ approaches to infinity, which implies $h$  is divergent. We label the corresponding  horizontal axis of the divergent point as $r_{0dh}$. The other is the final location of the curve,  where $h$ vanishes. Obviously, in this case $r_0\rightarrow r_h$.  The corresponding  horizontal axis of the critical  point is labeled as $r_{0ch}$.
In fact, for the plane symmetric black holes, \cite{Leichenauer:2014nxa} has obtained these results analytically. It was  found that at $r_{0dh}$, $\Xi_3$ diverges thus $h$ approaches to infinity. At  $r_{0ch}$, $h$ vanishes for both  $\Xi_1$ and $\Xi_2$ behave as $\log(r_h-r_0)$. Our results show that these conclusions are still valid  for the spherically symmetric black holes.
We also investigate the effect of the charge on the shift $h$. We can see that as the charge increases, both the values of the  divergent point  and critical point become smaller.
In addition, we find
for a fixed $r_0$, greater value of the charge corresponds to smaller shift $h$, which implies the charge delays the formation of the shock wave geometry.

\begin{figure} [h]
\centering
\includegraphics[scale=0.65]{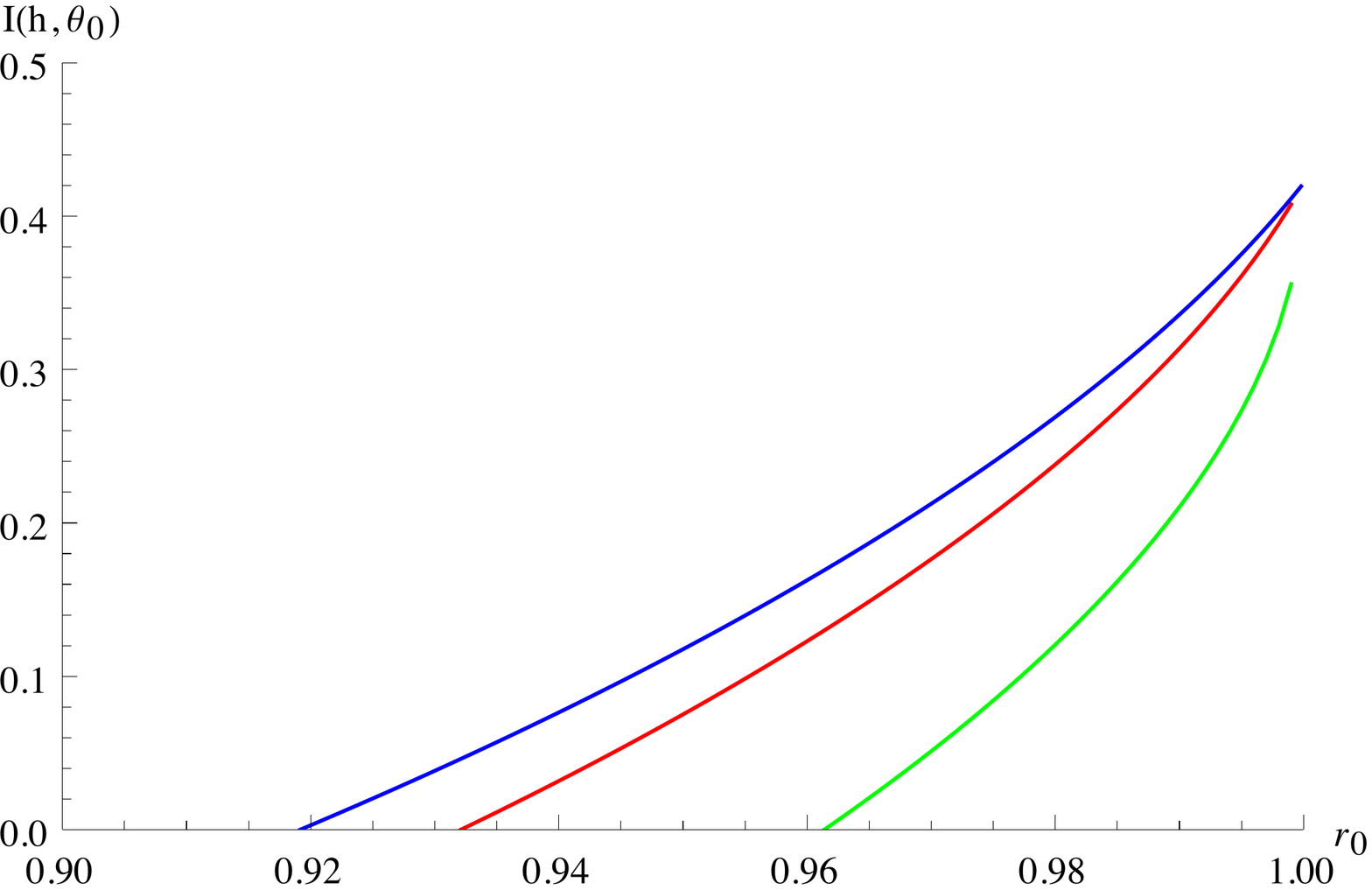}
 \caption{\small  Relation between  $I(h, \theta_0)$ and  $r_{0}$ for the case $r_{min} = 50, r_h=1$. The green line, red line, and blue line correspond to $Q=0.5, 0.52, 0.54$ respectively.  } \label{fig102}
\end{figure}

With  \Eq{ih}, we can get the relation between $I(h, \theta_0) $ and $r_0$,  which is  shown in  \Fig{fig102}.
 We can see that for a fixed charge, $I(h, \theta_0) $  increases as $r_0$ increases. Especially, there  is a critical value of  $r_0$, where  $I(h, \theta_0) $ vanishes.  We label the corresponding  horizontal axis of the critical  point as $r_{0ci}$. We also investigate the effect of the charge on the critical point $r_{0ci}$ and find that larger the value of the charge is, smaller the  value of $r_{0ci}$ will be.
 For a fixed value of $r_0$, the mutual correlation is bigger as the charge $Q$ becomes greater. It seems contradict with the statements in  section 3  where the mutual correlation decreases with respect to the charge. The readers should note that in section 3   there is no shake wave added in the background while there is. This observation indicates  that   the dynamical shock wave geometry   have dominant impact to the mutual correlation in the shock wave geometry.

\begin{figure} [h]
\centering
\includegraphics[scale=0.65]{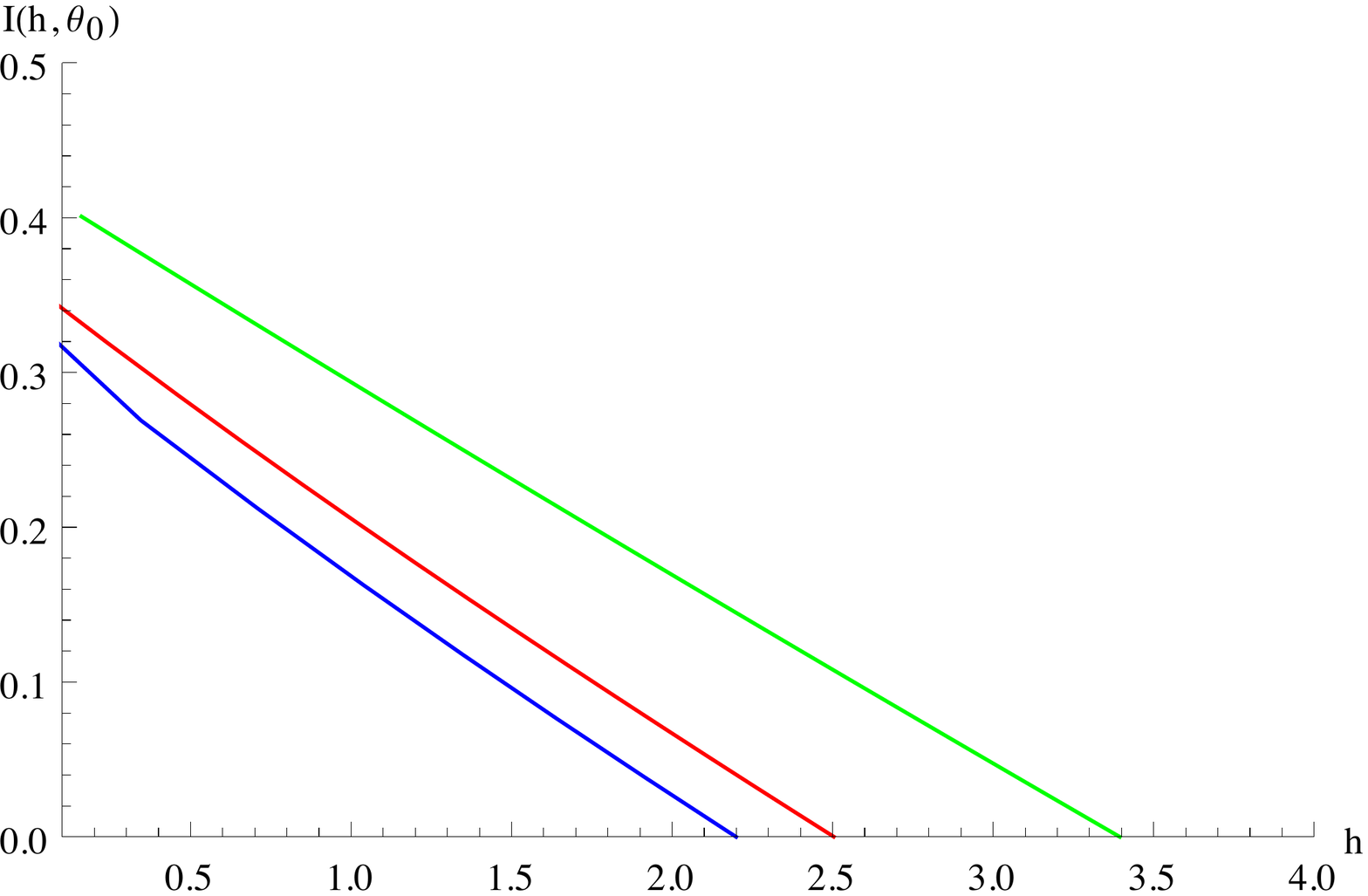}
 \caption{\small Relation between  $I(h, \theta_0)$ and  $h$  for the case $ \bar{r}=0.2, r_{min} = 50$. The green line, red line, and blue line correspond to $Q=0.5, 0.52, 0.54$ respectively.  } \label{fig11}
\end{figure}

 Having obtained the relation between $h$ and $r_0$  as well as $I(h, \theta_0) $ and $r_0$, we can obtain the relation between $I(h, \theta_0) $ and $h$, which is shown in \Fig{fig11}. It is obvious that as  $h$ increases,
  $I(h, \theta_0)$ decreases. There is also a critical value of $h$, labeled as $h_c$,  where $I(h, \theta_0)$ vanishes. With these observations, we can conclude  that the perturbation added at the left boundary will disrupt  the wormhole geometry, and as the  wormhole geometry grows to a critical value, the mutual correlation  vanishes for the left region and the right region is uncorrelated now.

For a fixed $h$ we also investigate the effect of the charge on the mutual correlation $I(h, \theta_0)$. Obviously, the larger the value of the charge is, the smaller the value of  the  mutual correlation $I(h, \theta_0)$ will be. This is similar to that of the  static case in section 3, for in this case, the effect of the charge is dominated. The effect of the charge on the  critical point  $h_c$ is also investigated. The larger the value of the charge is, the smaller the value of the horizontal coordinate of the critical point  $h_c$ will be. That is, in the shock wave geometry, the charge will prompt the correlated two quantum system on the boundary of the AdS spacetime to be uncorrelated.

\section{Conclusion and discussion}
Usually, one often uses the mutual information, defined by the holographic entanglement entropy, to probe the entanglement of two regions living on the boundary of the AdS black holes. In \cite{Leichenauer:2014nxa}, the author investigated the mutual information of the  Reissner Nordstr\"om  AdS black holes with and without shock wave geometry. For the static case,
 they found that for large boundary regions the mutual information is positive while for small ones it vanishes. In the shock wave background, they found that the mutual information is disrupted by the perturbation added at the boundary, and for large enough perturbation, the mutual information vanishes, which implies the left region and right region are uncorrelated.

  In this paper,
we  employed  the mutual correlation, which is defined  by the geodesic length,  to probe the correlation of two regions living on the boundary of the Reissner Nordstr\"om  AdS black holes.
We first investigated the mutual correlation  in the static background. We found that
as the size of the  boundary  region is large enough, the value of the mutual correlation is positive always, namely
the
two regions living on the boundary of the  AdS black holes are correlated. Our result implies that the mutual correlation has the same effect as that of the mutual information as  they are used to probe the correlation  of two regions.
  We also investigated the effect of the charge on the  mutual correlation and found it decreases as the charge increases. That is, the charge will destroy the correlation of correlated two regions.

By adding the perturbations into the bulk, we studied the dynamic mutual correlation in the shock wave geometry.  We found that as the added perturbation becomes greater, the shift of the horizon becomes larger, and
the mutual  correlation  decreases rapidly. Especially, there is a critical  value for the shift where the  mutual correlation
  vanishes as the perturbation is large enough. Obviously, our result is also the same as that probed by the mutual information in \cite{Leichenauer:2014nxa}.
   We also investigated the effect of the charge on the mutual  correlation  and found that
    the bigger the value of the charge is, the smaller the value of the mutual  correlation will to be. Namely, the charge will destroy the  correlation of  the correlated two regions, which  is the same as that in the static background.

 In \cite{Shenker:2013pqa}, it has been found that for a spin system, the two point functions and mutual information have a qualitatively
similar response to a perturbation of the thermofield double state.
Thus it is also interesting to use directly the  two point functions to probe the butterfly effect though it is more crude relatively compared with the mutual information and mutual correlation \cite{Shenker:2013pqa}.

\section*{Data Availability}

All the figures can be obtained by the corresponding equations and values
 of the parameter. We did not adopt other data.

\section*{Acknowledgements}
We are grateful to Hai-Qing Zhang for his instructive discussions. This
work is supported by the National Natural Science Foundation of China (Grant Nos. 11405016), and Basic Research Project of Science and Technology Committee of Chongqing(Grant No. cstc2016jcyja0364).




\end{document}